\documentclass[graybox]{svmult}
\usepackage{bm}

\usepackage{mathptmx}       
\usepackage{helvet}         
\usepackage{courier}        
\usepackage{type1cm}        
\usepackage{makeidx}         
\usepackage{graphicx}        
\usepackage{multicol}        
\usepackage[bottom]{footmisc}

\makeindex             

\begin{document}

\title*{Online Machine Learning in Big Data Streams}
\author{Andr\'as A. Bencz\'ur and Levente Kocsis and R\'obert P\'alovics} 
\institute{All authors \at Institute for Computer Science and Control, Hungarian Academy of Sciences (MTA SZTAKI), \email{\{benczur,\,kocsis,\,rpalovics\}@sztaki.mta.hu}}
%
%
\maketitle

\abstract{The area of online machine learning in big data streams covers algorithms that are (1) distributed and (2) work from data streams with only a limited possibility to store past data.
  The first requirement mostly concerns software architectures and efficient algorithms.  The second one also imposes nontrivial theoretical restrictions on the modeling methods: In the data stream model, older data is no longer available to revise earlier suboptimal modeling decisions as the fresh data arrives.
  In this article, we provide an overview of distributed software architectures and libraries as well as machine learning models for online learning.  We highlight the most important ideas for classification, regression, recommendation, and unsupervised modeling from streaming data, and we show how they are implemented in various distributed data stream processing systems.
  This article is a reference material and not a survey.  We do not attempt to be comprehensive in describing all existing methods and solutions; rather, we give pointers to the most important resources in the field.  All related sub-fields, online algorithms, online learning, and distributed data processing are hugely dominant in current research and development with conceptually new research results and software components emerging at the time of writing. In this article, we refer to several survey results, both for distributed data processing and for online machine learning.  Compared to past surveys, our article is different because we discuss recommender systems in extended detail.
}

\section{Introduction}

Big data analytics promise to deliver valuable business insights. However, this is difficult to realize using today's state-of-the-art technologies, given the flood of data generated from various sources. A few years ago, the term \textbf{fast data}~\cite{lam2012muppet} arose to capture the idea that \textbf{streams} of data are generated at very high rates and that these need to be analyzed quickly in order to arrive at actionable intelligence.  

Fast data can flood from network measurements, call records, web page visits, sensor readings, and so on~\cite{bifet2011data}.  The fact that such data arrives continuously in multiple, rapid, time-varying, possibly unpredictable, and unbounded streams appears to yield some fundamentally new research problems.
Examples of such applications include financial applications~\cite{zhu2002statstream}, network monitoring~\cite{babu2001continuous,abadi2003aurora}, security, sensor networks~\cite{de2016iot}, Twitter analysis~\cite{bifet2010sentiment}, and more~\cite{fontenla2013online}.

Traditional data processing assumes that data is available for multiple access, even if in some cases it resides on disk and can only be processed in larger chunks.
In this case, we say that the data is \textbf{at rest}, and we can perform \textbf{batch processing}.
Database systems, for example, store large collections of data and allow users to initiate queries and transactions.

Fast data, or \textbf{data in motion} is closely connected to and in certain cases used as a synonym of the \textbf{data stream} computational model~\cite{muthukrishnan2005data}.
In this model, data arrives continuously in a potentially infinite stream that has to be processed by a resource-constrained system.  The main restriction is that the main memory is small and can contain only a small portion of the stream, hence most of the data has to be immediately discarded after processing.

In one of the earliest papers that describe a system for data stream processing~\cite{abadi2003aurora}, the needs of monitoring applications are described.  The tasks relevant for monitoring applications differ from conventional processing of data at rest in that the software system must process and react to continuous input from multiple sources.  The authors introduce the data active, human passive model, in which the system permanently processes data to provide alerts for humans.

Needs and opportunities for \textbf{machine learning over fast data streams} are stimulated by a rapidly growing number of industrial, transactional, sensor and other applications~\cite{zliobaite2012next}.
The concept of online machine learning is summarized in one of the earliest overviews of the field~\cite{widmer1996learning}.  The principal task is to learn a concept incrementally by processing labeled training examples one at a time.  After each data instance, we can update the model, after which the instance is discarded.

In the data stream computational model, only a small portion of the data can be kept available for immediate analysis~\cite{henzinger1998computing}.  This has both algorithmic and statistical consequences for machine learning: Suboptimal decisions on earlier parts of the data may be difficult to unwind, and if needed, require low memory sampling and summarization procedures.  For data streaming applications, incremental or online learning fits best.

\begin{quote}
\textbf{Requirement 1}: Online learning updates its model after each data instance without access to all past data, hence the constraints of the data streaming computational model apply.
\end{quote}

Data streaming is not just a technical restriction on machine learning: Fast data is not just about processing power but also about fast semantics.
Large databases available for mining today have been gathered over months or years, and the underlying processes generating them have changed during this time, sometimes radically~\cite{hulten2001mining}.
In data analysis tasks, fundamental properties of the data may change quickly, which makes gradual manual model adjustment procedures inefficient and even infeasible~\cite{zliobaite2012next}.
Traditional, batch learners build static models from finite, static, identically distributed data sets.  By contrast, stream learners need to build models that evolve over time.  Processing will strongly depend on the order of examples generated from a continuous, non-stationary flow of data. Modeling is hence affected by potential concept drifts or changes in distribution~\cite{gama2013evaluating}.

\begin{quote}
\textbf{Requirement 2}: Adaptive machine learning models are needed to handle concept drift.
\end{quote}
  
As an additional consequence of Requirement 2, adaptive learning also affects the way evaluation is performed.  Potentially in every unit of time, the system may return predictions from different models, and we may receive too few predictions from a particular model to evaluate by traditional metrics.
Instead, we have to define error measures that we can minimize in a feedback system:  Predictions are made for a stream of objects one by one, and the correct answer is received immediately afterwards.  A discrepancy between the prediction and the observed value serves as feedback, which may immediately trigger modifications to the model~\cite{widmer1996learning}.

Finally, the third important aspect for online learning from big data  is algorithmic.  In order to cope with the volume of the data, processing has to be distributed.
Clusters of machines are hard to manage, and hardware failure must be mitigated in the case of an application running on thousands of servers. 
Map-Reduce~\cite{dean2008mapreduce} was the first programming abstraction designed to manage the cluster, provide fault tolerance, and ease software development and debugging.
While Map-Reduce is designed for batch processing, distributed data stream processing needs other solutions, such as communication between processing elements~\cite{neumeyer2010s4} via an interconnection topology~\cite{toshniwal2014storm}.
For an outlook, mostly batch distributed data mining solutions are surveyed in~\cite{fan2013mining}.

\begin{quote}
\textbf{Requirement 3}: Online learning from big data has to be implemented in a distributed stream processing architecture.
\end{quote}

Several surveys for online machine learning in general~\cite{gaber2007survey,bifet2011data,shalev2012online,fontenla2013online,gaber2014data} and subfields~\cite{widmer1996learning,tsymbal2004problem,cheng2008survey,mahdiraji2009clustering,quionero2009dataset,kavitha2010clustering,zliobaite2012next,aggarwal2013survey,silva2013data,gama2014survey} have appeared recently.  Our survey is different, first of all, in that we focus on three aspects: the data stream computational model, the adaptive methods for handling concept drift, and the distributed software architecture solutions for streaming.
We elaborate on systems for machine learning by distributed data stream processing.  In particular, we explore the idea of using Parameter Servers.
We focus on nonstatic environments and give no convergence theorems for performance on identically distributed streams.

To the best of our knowledge, this is the first survey on online machine learning with an extensive discussion of recommender systems. 
Recommenders are important as they give a clear, industry-relevant example of Requirement~2.  Note that in~\cite{zliobaite2012next} it is observed that adaptive learning models are still rarely deployed in industry.

This paper is organized as follows:  In Section~\ref{sect:systems} we give an overview of distributed software architectures for online machine learning.  Then in Sections~\ref{sect:supervised}--\ref{sect:additional} we list some of the most important online learning models for supervised classification, reinforcement learning, recommendation, unsupervised analysis, and concept drift mitigation.

\section{Processing Data Streams for Machine Learning}
\label{sect:systems}

Data-intensive applications often work in the data stream computational model~\cite{babcock2002models}, in which the data is transient:  Some or all of the input data is not available for random access from disk or memory.  The data elements in the stream arrive online and can be read at most once; in case of a failure, it is possible that the data elements cannot be read at all.
The system has no control over the order in which data elements arrive to be processed either within a data stream or across data streams.

The strongest constraint for processing data streams is the fact that once an element from a data stream has been processed, it has to be discarded or archived.  Only selected past data elements can be accessed by storing them in memory, which is typically small relative to the size of the data streams. 
Many of the usual data processing operations would need random access to the data~\cite{babcock2002models}:  For example, only a subset of SQL queries can be served from the data stream.
As surveyed in~\cite{muthukrishnan2005data}, data stream algorithms can tackle this constraint by a variety of strategies, including adaptive sampling in sliding windows, selecting representative distinct elements, and summarizing data in low-memory data structures, also known as sketches or synopses.  

When designing online machine learning algorithms, we have to take several algorithmic and statistical considerations into account.
The first problem we face is the restrictions of the computational model.  As we cannot store all the input, we cannot unwind a decision made on past data.  For example, we can use statistical tests to choose from competing hypotheses~\cite{domingos2000mining}, which give theoretical guarantees in the case of identically distributed data.

A second problem with real stream learning tasks is that data typically changes over time, for example, due to concept drifts~\cite{widmer1996learning}.  For changing distributions, we can even use the streaming computational model to our advantage: By permanent training, we can adapt to concept drift by overwriting model parameters based on insights from fresh data and thus forgetting the old distribution~\cite{frigo2017online}.  Possible means of concept drift adaptation, however, will depend on the task, the data, and the choice of the algorithm~\cite{widmer1996learning,klinkenberg2000detecting}.

In this section we give a brief overview of how different, mostly open source, software projects manage machine learning tasks over streaming data.  We note that very active, ongoing research in the field may make parts of our description obsolete very quickly.
First, in Section~\ref{sect:dspe} we provide a summary of the most important distributed data stream processing engines with active development at the time of writing.  Next, in Section~\ref{sect:ml-taxonomy} we give a taxonomy of the possible machine learning solutions with respect to support from streaming data and distributed processing.  Since operation on shared-nothing distributed architectures is a key requirement for big data processing solutions, we discuss the main machine learning parallelization strategies in Section~\ref{sect:parallelism}.  One particularly popular solution, the parameter server is also described in more detail.  Finally, in Section~\ref{sect:stream-ml} we list functionalities of stream learning libraries as of the time of writing.

\subsection{Data Stream Processing Engines}
\label{sect:dspe}

Distributed stream learning libraries are usually either part of \textbf{data stream processing engines (DSPEs)} or built on top of them through interfaces.  For a better understanding of the software architecture, we give an overview of DSPEs next.

Since the emergence of early systems such as Aurora~\cite{abadi2003aurora,arasu2003stream}, several DSPEs have been developed. For a survey of DSPEs, see~\cite{ranjan2014streaming}.
The main focus of the most recent, evolving DSPEs is to provide simplicity, scalability, stateful processing, and fault tolerance with fast recovery.  In terms of simplicity, the most important goal is to overcome the need to integrate the DSPE with a batch processing engine as in the so-called lambda architecture described, among other places, in~\cite{marz2015big,kiran2015lambda}.

In this section we focus on Apache Spark~\cite{zaharia2010spark} and Apache Flink~\cite{carbone2015apache}, since at the time of writing these DSPEs have the most active development for learning from streams.  Other systems usually provide machine learning functionalities by interfacing with SparkML, a Spark-based machine learning library, or SAMOA~\cite{morales2015samoa}, a stream learning library designed to work as a layer on top of general DSPEs.  We list the functionality of these systems in Section~\ref{sect:stream-ml}. 

Without attempting to be exhaustive, we list some other main DSPEs.  Note that this field is very active and several systems with large impact on both DSPEs and stream learning have already stopped development.  For example, several active projects have borrowed concepts from the \textbf{S4 project}~\cite{neumeyer2010s4}, which retired in 2014.
Proprietary systems are summarized in~\cite{gualtieri2013forrester}; it appears to be the case that commercial developers as of yet have no special focus on data stream processing, hence our main goal is to give an overview of the open source solutions.

\textbf{Storm}~\cite{toshniwal2014storm} relies on the concept of the connection topology of the processing elements.  The topology is allowed to contain cycles; however in case of a failure with cycles, neither the exactly-once nor the at-least-once processing condition can be enforced.  
\textbf{Samza}~\cite{noghabi2017samza} provides a design for fast recovery after failures independent of the state size.  As a new advantage, stateful streaming systems have the possibility to emulate batch operation, hence the need for the lambda architecture can be eliminated. 
\textbf{Beam}~\cite{akidau2015dataflow}, based on the Google Cloud dataflow model~\cite{chambers2010flumejava,akidau2013millwheel,akidau2015dataflow}, focuses on event time windows to process out-of-order data.  Beam can be connected to the deep learning framework Tensorflow~\cite{dean2012large} for batch machine learning.
Finally, while most of the above systems are distributed, we mention Esper~\cite{bernhardt2007esper} as a prominent single machine, in-memory DSPE.

One DSPE in the focus of this article, \textbf{Spark}~\cite{zaharia2010spark} treats stream processing as a sequence of small batch computations. Records in the stream are collected into micro-batches (by time), and a short-lived batch job is scheduled to process each micro-batch.  The advantage of this approach is its intuitive transition from batch to streaming with straightforward fault tolerance and interaction with batch programs.  The main disadvantage is higher latency due to the inherent scheduling overhead for the micro-batch.

Another DSPE of our choice, \textbf{Flink}~\cite{carbone2015apache} provides consistent managed state with exactly-once guarantees, while achieving high throughput and low latency, serving both batch and streaming tasks.
Flink handles the streams event by event in a true streaming fashion through the underlying streaming (dataflow) runtime model. While this provides more fine-grained access to the stream, it does not come with a throughput overhead due to various runtime optimizations such as buffering of output records. The advantages are very low processing latency and a natural stateful computational model.  The disadvantages are that fault tolerance and load balancing are more challenging to implement.  Flink is primarily for stream processing.  Flink batch tasks can be expressed by using loops in stream processing, as we will see in the next section.

\subsection{Taxonomy of Machine Learning Tools}
\label{sect:ml-taxonomy}

\begin{figure}
  \sidecaption
\includegraphics[width=7cm]{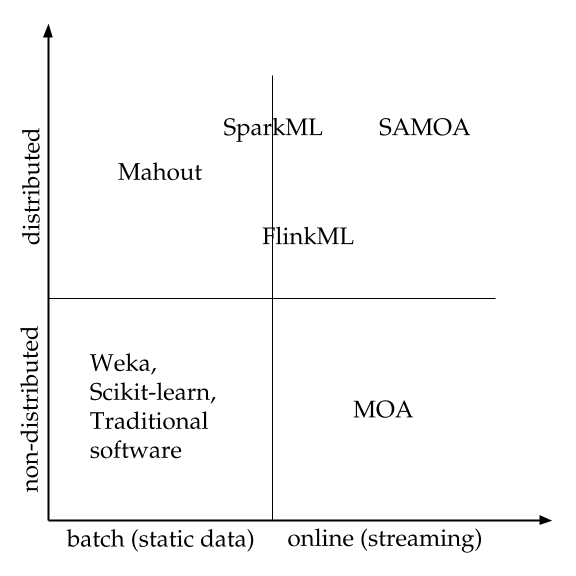}
\caption{Taxonomy of machine learning tools.}
\label{fig:taxonomy}
\end{figure}

In this article, we survey online machine learning tools for big data.   Our main focus are models, architectures, and software libraries that process streaming data over distributed, shared-nothing architectures.  As shown in Fig.~\ref{fig:taxonomy}, the two key distinguishing features of machine learning tools are whether they are distributed and whether they are based on static or streaming data.

As the least restrictive of the four quadrants in Fig.~\ref{fig:taxonomy}, \textbf{batch non-distributed} tools can implement any method from the other quadrants.  If scale permits, data streams can first be stored and then analyzed at rest, and distributed processing steps can be unfolded to run sequentially on a single-processor system.  Traditional machine learning tools, for example, R, Weka, and scikit-learn, fall into this quadrant.

\textbf{Distributed batch} machine learning systems~\cite{fan2013mining} typically implement algorithms by using the Map-Reduce principle~\cite{dean2008mapreduce}.  Perhaps the best-known system is Mahout~\cite{owen2011mahout} built on top of Hadoop~\cite{white2010hadoop}, but a very rich variety of solutions exists in this field.  GraphLab~\cite{low2012distributed}, withdrawn from the market in 2016, was another mostly batch tool that has also influenced online systems.

\textbf{Online learning} solutions cover algorithms that immediately build models after seeing a relatively small portion of the data.  By this requirement, we face the difficulty of not necessarily being able to undo a suboptimal decision made in an earlier stage, based on data that is no longer available for the algorithm.

The first library for online machine learning, MOA~\cite{bifet2010moa} collects a variety of models suitable for training online, most of which we describe in Sections~\ref{sect:supervised} and~\ref{sect:unsupervised}.  Based on MOA concepts, SAMOA~\cite{morales2015samoa} is a distributed framework---a special purpose DSPE---and library that provides distributed implementation for most MOA algorithms.  Online learning recommender systems, both distributed and non-distributed, are also described in~\cite{palovics2017tutorial}.  Another recent non-distributed online learning tool is described in~\cite{bifet2017extremely}.

For a \textbf{distributed online learning} software architecture, the underlying system needs to be a DSPE.  In addition to SAMOA, which can be considered a DSPE itself, most DSPEs of the previous section can be used for distributed online learning.  For example, Flink, Samza, and Storm implement interfaces to use SAMOA libraries.  On the other hand, a DSPE can also implement batch machine learning algorithms: For example, the machine learning library SparkML is mostly batch and FlinkML is partly batch.

Combined batch and online machine learning solutions are of high practical relevance.  We can train our models batch based on a precompiled sample, and then apply prediction for a live data stream.  DSPE solutions are progressing in this area.  One recent result, Clipper is a low-latency online prediction system~\cite{crankshaw2017clipper} with a model abstraction layer that makes it easy to serve pre-trained models based on a large variety of machine learning frameworks.

Another batch approach to learning from time-changing data is to repeatedly apply a traditional learner to a sliding window of examples:  As new examples arrive, they are inserted into the beginning of the window.  Next, a corresponding number of examples are removed from the end of the window, and the learner is reapplied, as in early research on concept drift in learning from continuous data~\cite{widmer1996learning,hulten2001mining}.  Finally, as a combination with online learning methods, the batch trained model can be incrementally updated by a streaming algorithm~\cite{frigo2017online}.

\subsection{Parallel Learning and the Parameter Server}
\label{sect:parallelism}

In order to design distributed modeling algorithms, we have to elaborate on parallelization strategies. 
\textbf{Horizontal} or \textbf{data parallel} systems partition the training data and build separate models on subsets that may eventually get merged.  Such a solution is applied, for example, for training XGBoost trees~\cite{chen2016xgboost}.  This approach, however, will typically depend on the partitioning and lead to heuristic or approximate solutions.

\textbf{Vertical parallel} or \textbf{model parallel training} solutions partition across attributes in the same data point, rather than partitioning the training data coming from the stream.  Each training point is split into its constituting attributes, and each attribute is sent to a different processing element.  For example, for linear models trained by gradient descent, coefficients can be stored and updated by accessing a distributed store~\cite{li2014scaling}.  As another example, the fitness of attributes for a split in a decision tree construction can be computed in parallel~\cite{morales2015samoa}.  Further examples such as Tensorflow~\cite{dean2012large}, Petuum~\cite{xing2015petuum}, and MXNet~\cite{chen2015mxnet} are also capable of model parallel training.  The drawback is that in order to achieve good performance, there must be sufficient inherent parallelism in the modeling approach. 

\begin{figure}
  \sidecaption
\includegraphics[width=7cm]{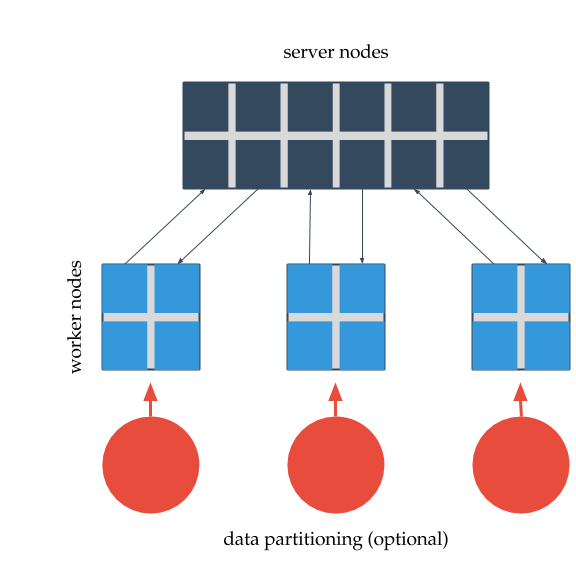}
\caption{The parameter server architecture for distributed machine learning.}
\label{fig:parameter-server}
\end{figure}

The \textbf{parameter server} introduced in~\cite{smola2010architecture} is a popular way to implement model parallel training, with several variants developed~\cite{ho2013more,li2013parameter,li2014scaling,li2014communication}, including an application for deep learning~\cite{dean2012large}.
The main idea is to simplify the development of distributed machine learning applications by allowing access to shared parameters as key--value pairs.  As shown in Fig.~\ref{fig:parameter-server}, the compute elements are split into two subsets: Parameters are distributed across a group of \textbf{server nodes}, while data processing and computation are performed at \textbf{worker nodes}. Any node can both push out its local parameters and pull parameters from remote nodes.  Parallel tasks are typically asynchronous, but the algorithm designer can be flexible in choosing a consistency model.

Various parameter server systems are summarized best in~\cite{li2014scaling}.  The first generation~\cite{smola2010architecture} uses distributed key-value stores such as memcached~\cite{fitzpatrick2004distributed} to store the parameters.  More recent solutions~\cite{dean2012large,li2013parameter} implement specific parameter access interfaces.  For example, as the first step towards a general platform,~\cite{ho2013more} design a table-based interface for stale synchronous operation, which they use to implement a wide variety of applications.

Most results~\cite{smola2010architecture,ho2013more,li2014scaling} describe parameter servers for variants of regression as well as unsupervised modeling by Latent Dirichlet Allocation (see Section~\ref{sect:unsupervised}).  Another popular use is classification by deep neural networks~\cite{dean2012large,li2013parameter}.
An implementation for Multiple Additive Regression Trees is given in~\cite{zhou2017psmart}.

Recommendation algorithms can also be implemented by the parameter server principle.
Batch implementations of asynchronous distributed stochastic gradient descent recommender algorithms are given in~\cite{ho2013more,schelter2014factorbird}.  
A recent comparison of distributed recommenders, including an online one based on a Flink parameter server, is given in~\cite{palovics2017tutorial}.

The parameter server idea fits the data stream as well as the batch computational models.  For example, if it suffices to read the data only once and in input order for gradient descent, a batch parameter server such as in~\cite{li2014scaling} immediately yields a streaming algorithm. Yet, most applications are for batch modeling only~\cite{smola2010architecture,ho2013more,li2014scaling,schelter2014factorbird,zhou2017psmart}.
As examples of parameter servers for online learning, applications for reinforcement learning can be found in~\cite{nair2015massively} and for recommenders in~\cite{palovics2017tutorial}.

\subsection{Stream Learning Libraries}
\label{sect:stream-ml}

We provide an overview of the present machine learning functionalities of the main open source engines.
Note that all these engines are under active development with a very large number of components already available as research prototypes or pull requests.  For this reason, we only want to give a representative overview of the main components.

\subsubsection{MOA and SAMOA}

SAMOA~\cite{morales2015samoa}, a distributed online learning library, is based on the concepts of MOA~\cite{bifet2010moa}, a single machine library.
SAMOA provides model parallel implementations of specific algorithms by using the concepts of Storm~\cite{toshniwal2014storm}: Processing elements are connected in a loop-free topology, which connects the various pieces of user code.  SAMOA can run on top of a DSPE with a flexible interface connection.  SAMOA connectors are implemented for all active DSPEs in Section~\ref{sect:dspe}.

For classification, SAMOA \cite{de2013samoa,morales2015samoa} provides the Vertical Hoeffding Tree (VHT), a distributed version of a streaming decision tree~\cite{domingos2000mining} (Section~\ref{sect:trees}). For clustering, it includes an algorithm based on CluStream~\cite{aggarwal2003framework} (Section~\ref{sect:unsupervised}).   The library also includes meta-algorithms such as bagging and boosting (Section~\ref{sect:ensemble}). 

\subsubsection{Apache Spark}

Spark has a very rich set of batch machine learning functionalities, including linear, tree, Support Vector Machine, and neural network models for classification and regression, ensemble methods as well as explicit and implicit alternating least squares for recommender systems, and many more listed at \url{https://spark.apache.org/mllib/}.  However, the only streaming algorithm in Spark MLLib is a linear model with ongoing work regarding online recommender and Latent Dirichlet Allocation prototypes.  Spark has no SAMOA connector yet.

Spark has several parameter server implementations, most of which are batch only.  We mention two projects with recent activity. 
Glint is a parameter server based Latent Dirichlet Allocation (Section~\ref{sect:unsupervised}) implementation, which is described in~\cite{jagerman2017computing}.
Angel is a general parameter server~\cite{jiang2017angel} that has implementations for logistic regression,  SVM (Section~\ref{sect:linear}), matrix factorization (Section~\ref{sec:mf}), Latent Dirichlet Allocation, and more.  Angel supports synchronous, asynchronous, and stale synchronous processing~\cite{jiang2017heterogeneity}.


\subsubsection{Apache Flink}

Flink has a loosely integrated set of machine learning components, most of which are collected at \url{https://github.com/FlinkML}.  In addition to the SAMOA connector at \url{https://github.com/apache/incubator-samoa/tree/master/samoa-flink}, Flink has a true streaming parameter server implementation at \url{https://github.com/FlinkML/flink-parameter-server}, which includes a Passive Aggressive classifier (Section~\ref{sect:linear}) and gradient descent matrix factorization (Section~\ref{sec:mf}).  Finally, Flink can serve online predictions by models trained on any system supporting the PMML standard~\cite{grossman2002data}, using the JPMML library at \url{https://github.com/FlinkML/flink-jpmml}.

In Flink, the parameter server is implemented as part of the data stream API at \url{https://github.com/FlinkML/flink-parameter-server}.  Since the communication between workers and servers is two-way, the implementation involves loops in stream processing.  As mentioned in Section~\ref{sect:dspe}, exactly-once processing and fault tolerance is conceptually difficult, and implementation is not yet complete as of the time of writing~\cite{carbone2017state}.

\section{Classification and Regression}
\label{sect:supervised}

The goal of classification and regression is to predict the unknown label of a data instance based on its attributes or variables.  The label is discrete for classification, and continuous for regression.
Classification and regression in batch settings are well established, with several textbooks available in data mining~\cite{pang2006introduction} and machine learning~\cite{friedman2001elements}.

For classification and regression over streaming data, the first two of the requirements in the Introduction play a key role.  By the first requirement, data is transient, since it does not fit in main memory, hence algorithms that iterate over the entire data multiple times are ruled out.  By the second requirement, we cannot assume that the examples are independent, identically distributed, and generated from a stationary distribution, hence different predictions by different models have to be applied and evaluated at different times. 
Special modeling tools are required to meet the two challenges, and known evaluation and comparison methods are not convenient yet~\cite{gama2009issues}.

An important, and perhaps the oldest, application of online learning is single trial EEG classification, in which the system learns from the online feedback of the experimental subjects~\cite{obermaier2001hidden}.  These early experiments differ from the approach in this section in that the performance was only measured at the end of the experiments, and no systematic analysis of the classifier performance in time was conducted.

This section is organized as follows:  First, we describe the difficulties and possibilities of evaluating online learning methods in Section~\ref{sect:classifier-eval}.
We cover the most important online classification and regression methods in the next subsections.  We discuss the main linear models in Section~\ref{sect:linear}, tree-based methods in Section~\ref{sect:trees}, classifier ensembles in Section~\ref{sect:ensemble},  Bayes models in Section~\ref{sect:bayes}, and finally, neural networks in Section~\ref{sect:neuralnet}. 
Other methods such as \textbf{nearest neighbor}~\cite{law2005adaptive} are known as well.
Our list of online classification methods is not comprehensive; for an extended list, see the recent survey~\cite{gaber2007survey}.

\subsection{Evaluation}
\label{sect:classifier-eval}

In an infinite data stream, data available for training and testing is potentially infinite.  Hence holdout methods for selecting an independent test set seem viable at first glance, and are used in a large number of online machine learning research results.  However, for online evaluation, we have no control over the order in which the data is processed, and the order is not independent of the data distribution. Since the distribution generating examples and the decision models evolve over time, cross-validation and other sampling strategies are not applicable~\cite{gama2009issues}.

Most studies of online learning determine overall loss as a sum of losses experienced by individual training examples.
Based on this so-called predictive sequential, abbreviated as \textbf{prequential} method~\cite{dawid1984present}, we define \textbf{online training and evaluation} in the following steps:
\begin{enumerate}
\item Based on the next unlabeled instance in the stream, we cast a prediction.
\item As soon as the true label of this instance becomes available, we assess the prediction error.
\item We update the model with the most recently observed error calculated by comparing the predicted and the true labels before proceeding with the next item of the data stream.
\end{enumerate}
Prequential error is known to be a pessimistic estimator, since it may be strongly influenced by the initial part of the sequence, when only a few examples have been processed, and the model quality is low.  The effect of the beginning of the stream can be mitigated, for example, by forgetting mechanisms~\cite{gama2009issues,gama2013evaluating}.

A further issue in online evaluation is that in reality, labels may arrive with delay~\cite{zliobaite2012next}.  The majority of adaptive learning algorithms require true labels to be available immediately after casting the prediction.  If true labels are delayed, we have to join two streams with time delay, one for the variables and one for the labels, which can make the implementation of the prequential evaluation scheme computationally challenging.

Error metrics that can be defined for individual data points can be applied for prequential evaluation.  For example, the definition of mean squared error, a popular metric for regression is
\begin{equation}
  MSE = \frac{1}{N} \displaystyle\sum_{i=1}^N  ( \hat{y_i} - y_i ) ^ 2,
\label{eq:mse}
\end{equation}
where $y_i$ is the actual and $\hat{y_i}$ is the predicted class label for data point $i$, and $N$ is the current size of the data stream.  Accuracy and error rate can be computed by a similar averaging formula.

For certain common metrics such as precision, recall, and true and false positive rates, the definition will involve the changing size of the set of  positive, negative, or all instances, which makes the metrics difficult to interpret in a very long stream of inhomogeneous data.  Although one definition of ROC~AUC~\cite{fogarty05roc} is based on true and false positive rates, its online interpretation is described in~\cite{zhao2011online}.  

\subsection{Linear Models}
\label{sect:linear}

Linear models in online machine learning date back to the perceptron algorithm~\cite{rosenblatt1958perceptron}.
The \textbf{perceptron learning} algorithm learns label $y$ of $d$-dimensional input vector $\mathbf{x}$ as a linear combination of the coordinates,
\begin{equation}
  \hat{y} = \mathbf{w} \cdot \mathbf{x}.
  \label{eq:perceptron}
\end{equation}
The prediction mechanism is based on an $n$-dimensional hyperplane of direction $\mathbf{w}$, which divides the instance space into two half-spaces. The margin of an example, $y \cdot \mathbf{w} \cdot \mathbf{x}$, is a signed value proportional to the distance between the instance and the hyperplane.  If the margin is positive, the instance is correctly classified; otherwise, it is incorrectly classified.

The perceptron can be trained by gradient descent for hinge loss as target function.  If labels $y$ take the values $\pm1$ and prediction $\hat{y}$ is defined by equation~(\ref{eq:perceptron}), hinge loss is equal to
\begin{equation}
  \ell(\mathbf{w}; (\mathbf{x}, y)) = \cases{
    0& if $y \cdot \hat{y}\ge 1$; \cr
    1 - y \cdot \hat{y} & otherwise,}
  \label{eq:hingeloss}
\end{equation}
from which the gradient can be computed as follows:  If prediction $\hat{y}$ has the correct sign, and its absolute value is at least 1, that is, hinge loss is 0, then there is no change; the algorithm is passive.  Otherwise, the gradient for $\mathbf{w}$ is $-\mathbf{x} \cdot \hat{y}$, and the update rule for learning rate $\eta$ is
\begin{equation}
   \mathbf{w} \leftarrow \mathbf{w} + \eta \cdot \mathbf{x} \cdot \hat{y} \mbox{\quad if } y \cdot \hat{y} < 1.
\end{equation}
The online gradient descent algorithm simply applies the above step to training examples in order~\cite{langford2009sparse}.  By contrast, the batch gradient descent algorithm reads the input multiple times and usually repeatedly optimizes coefficients for the same instance.  Batch gradient descent can be emulated by an online algorithm: We go through training examples one by one in an online fashion and repeat multiple times over the training data~\cite{cesa2008improved}.

Based on the general idea of perceptron learning, several online linear models have been proposed; for a detailed overview, see~\cite{fontenla2013online}.
The \textbf{Passive Aggressive (PA) classifier}~\cite{crammer2006online} is a popular online linear model that works well in practice, for example, applied as the Gmail spam filter~\cite{aberdeen2010learning}.
The main goal of the PA algorithm is to improve the convergence properties of the perceptron algorithm, the simplest numerical optimization procedure.  Several improved online optimization procedures were proposed prior to PA: Kivinen and Warmuth~\cite{kivinen1997exponentiated} give an overview of numerous earlier additive and multiplicative online algorithms, all of which are solvable by gradient descent~\cite{li2002relaxed,gentile2001new,crammer2003ultraconservative,kivinen2004online}. 
Based on the experiments in~\cite{crammer2003ultraconservative,crammer2006online}, the Passive Aggressive algorithm outperforms most of the earlier online linear classification methods.
Perceptron learning and most of its successors apply both to classification and regression; in other words, the range of the actual label $y$ can be both binary and continuous.

The PA algorithm solves a constrained optimization problem: We would like the new classifier to remain as close as possible to the current one while achieving at least a unit margin on the most recent example.  We use the Euclidean distance of the classifiers, $||\mathbf{w} - \mathbf{w}'||^2$. PA optimizes for hinge loss, with the goal to keep the distance minimal and the value of the margin at least 1, that is, to solve the optimization problem
\begin{equation}
  \mathbf{w} \leftarrow \mbox{argmin}_{\mathbf{w}'} ||\mathbf{w} - \mathbf{w}'||^2 \mbox{\quad s.t.\quad}
  \ell(\mathbf{w}; (\mathbf{x}, y)) = 0.
  \label{eq:pa-optimization}
\end{equation}
The algorithm is passive if loss is 0, and otherwise aggressive, since it enforces zero loss.
As shown in~\cite{crammer2006online}, the solution of the optimization problem yields the update rule
\begin{equation}
  \mathbf{w} \leftarrow \mathbf{w} + \ell(\mathbf{w}; (\mathbf{x}, y)) \cdot y \cdot \mathbf{x} / ||\mathbf{x}||^2.
\end{equation}
In~\cite{crammer2006online}, variants of PA are described that introduce a slack variable $\xi \ge 0$ in equation~(\ref{eq:pa-optimization}) such that the margin must stay below $\xi$. Adding constant times $\xi$ or $\xi^2$ to the minimization target in equation~(\ref{eq:pa-optimization}) leads to similar optimization problems.  In the same paper, multi-class, cost-sensitive, and regression variants are described as well.

Although the class of linear predictors may seem restrictive, the pioneering work of Vapnik~\cite{vapnik1998statistical} and colleagues demonstrates that by using kernels one can employ highly nonlinear predictors as well.  The \textbf{Support Vector Machine (SVM)} learns an optimal separating hyperplane $\mathbf{w}^*$ in a certain high-dimensional mapped space defined by the mapping $\varphi (\mathbf{x})$ over  training vectors $\mathbf{x}$.  Like with perceptron learning, the prediction has the form $\hat{y}=\mathbf{w}^* \cdot \varphi(\mathbf{x})$; however, $\mathbf{w}^*$ may be of very high dimensionality.

Potential problems of the very high dimensional mapped space are eliminated by the \textbf{kernel trick}. In the optimization procedure, $\mathbf{w}^*$ turns out to be the combination of the so-called support vectors, the subset $SV$ of training instances $\mathbf{x}_i$ for $i \in SV$ that maximize the margin.  We can obtain the parameters
\begin{equation}
  \mathbf{w}^* = \sum_{i \in SV} \alpha_i y_i \varphi(\mathbf{x}_i),
  \label{eq:kerneltrick}
\end{equation}
where $y_i$ are the labels and $\alpha_i$ can be found by maximizing the margin. By equation~(\ref{eq:kerneltrick}), we can reduce the prediction to computing inner products in the mapped space:
\begin{equation}
  \mathbf{w}^* \cdot \varphi(\mathbf{x}) =
  \sum_{i \in SV} \alpha_i y_i \varphi(\mathbf{x}_i) \cdot \varphi(\mathbf{x}),
\end{equation}
The main goal of online SVM is to maintain a set of support vectors and corresponding multipliers within the limits of available memory.  Whenever an online SVM learner decides to add new support vectors, others need to be discarded first, and $\alpha_i$ need to be updated.

To decide which support vector to discard, the definition of the span and the S-span of the support vectors defined by Vapnik~\cite{vapnik2000bounds} can be used.  The span of a support vector is the minimum distance of the vector from a certain set defined by all others.   The S-span of a set of support vectors is the maximum span among them.

An online SVM method to update the set of support vectors from the data stream is proposed in~\cite{agarwal2008kernel}, namely, maintaining a set of support vectors and multipliers that fit into the memory limit.  For a misclassified new instance, the algorithm measures the S-span of all possible ways of replacing one old support vector with the new instance, and selects the best one.  Multipliers are updated by the incremental learning procedure of~\cite{cauwenberghs2001incremental}.

Several other, similar online SVM optimization methods are known, for example~\cite{crammer2003ultraconservative,crammer2004online,bordes2005huller,bordes2005fast}.
Online learning algorithms were also suggested as fast alternatives to SVM in~\cite{freund1999large}, based on the online algorithm for calculating the maximum margin linear model of~\cite{frie1998kernel}.
Online gradient descent optimizers for linear models often work with kernels as well~\cite{kivinen2004online}.

\subsection{Decision and Regression Trees}
\label{sect:trees}

In a decision or regression tree, each internal node corresponds to a test on an attribute, while leaves contain predictors for classification or regression.  To build a tree, decision tree induction algorithms iterate through each attribute and compute an information theoretic function such as entropy or Gini index for classification and variance for regression~\cite{pang2006introduction}.

Online tree induction algorithms face the difficulty that the recursive tree construction steps cannot read past data.  After a split decision is made, batch algorithms partition all data into child nodes and compute the required information theoretic function of the attributes separately in each child node.  Online algorithms cannot partition past data; instead, they take advantage of the potentially infinite data and use fresh instances only in the newly created nodes.

Another problem with online tree construction is that a split decision has to be made at a certain point in time, without seeing future data.  A popular solution is to use the Hoeffding criterion for a statistical guarantee that the selected split is optimal for future data as well. 
In the so-called Hoeffding Tree or Very Fast Decision Tree (VFDT)~\cite{domingos2000mining}, attribute information theoretic functions are maintained over the stream.  If the Hoeffding criterion is met, a split decision is made, and the attribute statistics over the new child nodes are computed based on the new data from the stream.  Similar methods for regression trees are described in~\cite{alberg2012knowledge,ikonomovska2015online}.  For online vertical parallel distributed Hoeffding Trees, see~\cite{kourtellis2016vht}.

One problem in decision tree construction on streaming data is the  cost of maintaining attribute statistics for many-valued attributes.  For such attributes, a low granularity histogram has to be maintained.  In~\cite{jin2003efficient}, a method is described that partitions the range of a numerical attribute into intervals and uses statistical tests to prune these intervals.

Another difficulty is caused by nonstationary data, since each new child node is processed based on a new portion of data from the stream.  Decision trees over evolving streams are considered in~\cite{bifet2009adaptive,bifet2010adaptive,bifet2017extremely}.  We give an overview of general methods for nonstationary data in Section~\ref{sect:drift}.

\subsection{Ensemble Methods}
\label{sect:ensemble}

Ensemble methods build multiple, potentially different models on potentially different subsets of instances and attributes~\cite{pang2006introduction}.
The simplest example is \textbf{bagging} where several base models are trained based on samples with replacement.  Sampling with replacement can be simulated online~\cite{oza2005online,bifet2010leveraging}. 
A special ensemble technique, the online random forest algorithm is described in~\cite{denil2013consistency}.

A highly successful ensemble technique is \textbf{boosting}, in which we generate a sequence of base models by incorporating the prediction error of the previous models when constructing the next.  For example, in AdaBoost, an algorithm designed for online learning in its first description~\cite{freund1995desicion}, the next classifier is trained by weighting instances based on the function of the error of previous classifiers.  In gradient boosting~\cite{chen2016xgboost}, the next model is trained on the residual, that is, the difference of the training label and the continuous predicted label of the previous classifiers.

For online boosting, the difference compared to batch boosting is that the performance of a base model cannot be observed on the entire training set, only on the examples seen earlier.  Like with online decision tree induction, decisions need to be taken based only on part of the training data.  Various online boosting algorithms are described in~\cite{oza2005online,chen2012online,beygelzimer2015optimal}, based on the ideas of parallel boosting via approximation~\cite{fan1999application,palit2012scalable,lazarevic2002boosting}.
For gradient boosted trees~\cite{chen2016xgboost}, a recent, most successful classification method, the online version is described in~\cite{vasiloudis2017boostvht}.

\subsection{Bayes Models}
\label{sect:bayes}

Bayesian networks were one of the earliest applications for online learning~\cite{friedman1997sequential}, with the first methods mostly using mini-batch updates~\cite{buntine1991theory}. 
Bayesian learning methods maintain conditional probability tables $P(\mathbf{x}|y)$ by counting, where $\mathbf{x}$ is a feature vector and $y$ is its label.  Considering the conditional probability tables and the class distribution as priors, the predicted class of an instance will be the one that maximizes the posterior probability computed by the Bayes rule~\cite{pang2006introduction}.

The simplest, Naive Bayes model makes the ``naive'' assumption that each input variable is conditionally independent given the class label~\cite{duda2012pattern}.  While this model works surprisingly well~\cite{domingos1997optimality}, a weaker assumption is needed in Bayesian networks~\cite{pearl2014probabilistic}, in which we represent each variable with a node in a directed acyclic graph. Rather than assuming independence naively, we assume that each variable is conditionally independent given its parents in the graph. 

For online learning, it is easy to update the conditional probabilities both for Naive Bayes and for Bayesian networks~\cite{friedman1997sequential}, provided that they fit into internal memory.  Methods for updating the network structure online are described, for example, in~\cite{friedman1997sequential,chen2001approach}.

\subsection{Neural Networks}
\label{sect:neuralnet}

Neural networks and deep learning have shown great promise in many practical applications ranging from speech recognition~\cite{hinton2012deep} and visual object recognition~\cite{krizhevsky2012imagenet} to text processing~\cite{collobert2008unified}.
One of the earliest applications of online trained neural networks is EEG classification~\cite{haselsteiner2000using}.

Gradient descent is perhaps the most commonly used optimization procedure for training neural networks~\cite{lecun1998efficient}, which naturally leads to online learning algorithms~\cite{juang1998online} as well. 
The traditional formulation of gradient descent is impractical for very large neural networks.  A scalable online distributed version, Downpour SGD~\cite{dean2012large}, uses asynchronous parameter updates in conjunction with a parameter server (Section~\ref{sect:parallelism}).

\section{Reinforcement Learning}

Reinforcement learning is an area of machine learning concerned with agents taking
actions in an environment with the aim of maximizing some cumulative
reward. It is different from supervised learning in that the
environment does not provide a target behavior, only rewards
depending on the actions taken. 

The environment is typically assumed to be a Markov Decision Process
(MDP). Formally, we assume a set of states, $S$, a set of actions,
$A$, a transition probability function $P(s,a,s')$ denoting the
probability of reaching state $s'$ after taking action $a$ in state
$s$, and a reward function $R(s,a)$ denoting the immediate reward
after taking action $a$ in state $s$.  

While there is a wide range of reinforcement learning algorithms (see,
e.g., \cite{sutton1998reinforcement}), we focus here on algorithms that fit the streaming model and
(possibly) deal with nonstationary environments. The streaming model
of reinforcement learning is constrained not only by a continuous flow
of input data, but also by a continuous requisite to take actions.


\subsection{Algorithms for Stationary Environments}

Most reinforcement learning algorithms estimate the value of feasible
actions and build a policy based on that value (e.g., by choosing the
actions with the highest estimates with some additional exploration). An
alternative to value prediction methods are policy gradient methods
that update a parameterized policy depending on the performance. 

\subsubsection{Value Prediction}

The value of a state is the expected cumulative reward starting from a
given state and following a particular policy. In a similar way, the action
value is the expected reward starting from a given state with a
particular action. 


Value prediction methods estimate the value of the state or the value
of the actions in particular states. In the former case, to build a
policy from the estimated values, an additional transition
model is needed as well. Such a model is provided for some domains (e.g., by the
rules of a game), but in many cases the transition model needs to be
learned as well. Action-values can be used directly for constructing a
policy without the need for a model.

Temporal-difference (TD) learning learns the state value estimate $V(s)$ by
the following update rule after each state transition
$(S_t,S_{t+1})$:
$$V(S_t) \leftarrow (1-\alpha) V(S_t) + \alpha (R_t + \gamma
V(S_{t+1})),$$
where $\alpha$ is a step-size, and $\gamma$ is the discount factor.
TD learning was used in one of the first breakthroughs for reinforcement
learning, that is Tesauro's backgammon program \cite{tesauro1995td}. 

The best-known action-value prediction algorithm is Q-learning
\cite{watkins1992q}. For each occurrence of a transition
$(S_t,A_t,S_{t+1})$, the algorithm updates the action-value $Q(S_t,A_t)$ by
$$Q(S_t,A_t) \leftarrow (1-\alpha) Q(S_t,A_t) + \alpha (R_t + \gamma
\max_a Q(S_{t+1}, a)).$$
Q-learning using deep neural network to approximate the action-values
has been successfully applied to playing some Atari games at human
expert level \cite{mnih2015human}. 

Another algorithm that learns action-values is Sarsa \cite{sutton1996generalization}.  For each sequence $S_t, A_t, S_{t+1}, A_{t+1}$, the algorithm updates its estimates by
$$Q(S_t,A_t) \leftarrow (1-\alpha) Q(S_t,A_t) + \alpha (R_t + \gamma
Q(S_{t+1}, A_{t+1})).$$
Sarsa was successfully used by \cite{ipek2008self} for optimizing a DRAM
memory controller. 

The value prediction algorithms above were described with update rules
for a tabular representation. In most cases, function approximation is
used, and the update rules rely on a gradient step. Online enhancements
of gradient descent as well as eligibility traces
\cite{sutton1998reinforcement} can be applied to all variants. 

\subsubsection{Policy Gradient}

While using value functions is more widespread, it is also possible to use a parameterized
policy without relying on such functions. Parameterized policies are
typically optimized by gradient ascent with respect to the performance of the
policy. 

A policy gradient algorithm, the REINFORCE algorithm \cite{williams1992simple} was used to
optimize policy in a Go playing program that outperforms the best human players~\cite{silver2016mastering}.  

\subsection{Algorithms for Nonstationary Environments}

Most reinforcement learning algorithms, including those discussed in
the previous section, assume that the environment does not change over
time. While incremental algorithms such as Q-learning can adapt well
to nonstationary environments, it may be necessary to devise more
explicit exploration strategies that can cope with changes, for example, in
reward distribution.   

A special case of reinforcement learning is the multi-armed bandit
problem. In this case, the agent repeatedly selects an action from $K$
possible choices, obtaining a reward after each choice. This problem
retains the notion of reward; however, there are no states and
consequently no state-transitions. In the non-stochastic variant
\cite{auer2002nonstochastic}, the distribution of the rewards may
change over time arbitrarily. Standard algorithms for this problem are
Exp3 and its variants \cite{auer2002nonstochastic}, which rely
on an exponential selection algorithm, including some exploration terms
as well. Contextual bandits extend the bandit setting with the notion
of state (or context); however, state transitions are still
missing. This framework was used for instance in
\cite{li2010contextual} to select personalized new stories. 
We note that the distinguishing feature of recommendation in a bandit
setting is that the user can provide feedback only on the recommended
items. 


\section{Recommender Systems}

Recommender systems \cite{ricci2011introduction} serve to predict user preferences regarding items such as music tracks (Spotify),  movies (Netflix), products, books (Amazon), blogs, or microblogs (Twitter), as well as content on friends' and personal news feeds (Facebook).

Recommender systems can be categorized by the type of information they infer about users and  items.
Collaborative filtering \cite{amazon-recommender,sarwar01item} builds models of past user-item interactions such as clicks, views, purchases, or ratings, while content-based filtering \cite{lops2011content} recommends items that are similar in content, for example, share phrases in their text description.
Context-aware recommenders~\cite{adomavicius2011context} use additional information on the user and the interaction, for example user location and weather conditions. 
Recent events in a user session~\cite{koenigstein2013towards} serve as a special context.

A milestone in the research of recommendation algorithms, the Netflix Prize Competition~\cite{netflix-prize} had high impact on research directions.
The target of the contest was based on the one to five star ratings given by users, with one part of the data  used for model training and the other for evaluation.
As an impact of the competition, tasks now termed batch rating prediction were dominating research results.

Recommendation models rely on the feedback provided by the user, which can be \textbf{explicit}, such as one to five star movie ratings on Netflix \cite{adhikari2012unreeling}.
However, most recommendation tasks are \textbf{implicit}, as the user provides no like or dislike information.  Implicit feedback can be available in the form of time elapsed viewing an item or listening to a song, or in many cases, solely as a click or some other form of user interaction.  In~\cite{pilaszy2015neighbor}, the authors claim that 99\% of recommendation industry tasks are implicit.

As a main difference between recommendation and classification, classifiers usually work independently of the event whose outcome they predict. Recommender systems, on the other hand, may directly influence observations: They present a ranked top list of items~\cite{deshpande2004item}, and the user can only provide feedback for the items on the list.
Moreover, real systems process data streams where users request one or a few items at a time and get exposed to new information that may change their needs and taste when they return to the service next time.  Furthermore, an online trained model may change and return completely different lists for the same user even for interactions very close in time.

By the above considerations, real recommender applications fall in the category of top item recommendation by online learning for implicit user feedback, a task that has received less attention in research so far.  In this section, we show the main differences in evaluating such systems compared to both classifiers and batch systems, as well as describe the main data stream recommender algorithms.

Online recommenders seem more restricted than those that can iterate over the data set several times, and one could expect inferior quality from the online methods.
By contrast, in \cite{palovics2014exploiting,frigo2017online}, surprisingly strong performance of online methods is measured.

As an early time-aware recommender system example, the item-based nearest neighbor~\cite{sarwar01item} can be extended with time-decay \cite{ding2005time}.
Most of the early models, however, are time-consuming to compute, difficult to update from a data stream, and hence need periodical batch training.
Probably the first result in this area, the idea of processing transactions in chronological order to incrementally train a recommendation model first appeared in~\cite[Section~3.5]{takacs2009scalable}.
Streaming gradient descent matrix factorization methods were also proposed in~\cite{isaacman2011distributed,ali2011parallel}, who use Netflix and Movielens data and evaluate by RMSE.

The difficulty of evaluating streaming recommenders was first mentioned in \cite{lathia2009temporal}, although the authors evaluated models by offline training and testing split.  Ideas for online evaluation metrics appeared first in \cite{palovics2013temporal,vinagre2014evaluation,palovics2014exploiting}.  In \cite{vinagre2014evaluation}, incremental algorithms are evaluated using recall.  In~\cite{palovics2014exploiting}, recall is shown to have undesirable properties, and other metrics for evaluating online learning recommenders are proposed.

Finally, we note that batch distributed recommender systems were surveyed in~\cite{karydi2016parallel}.

\subsection{Prequential (Online) Evaluation for Recommenders}

To train and evaluate a time-sensitive or online learning recommender, we can use the prequential or online evaluation framework that we defined for classifier evaluation.  As seen in Fig.~\ref{fig:online}, online evaluation for a recommender system includes the following steps:
\begin{figure}
\includegraphics[width=12cm]{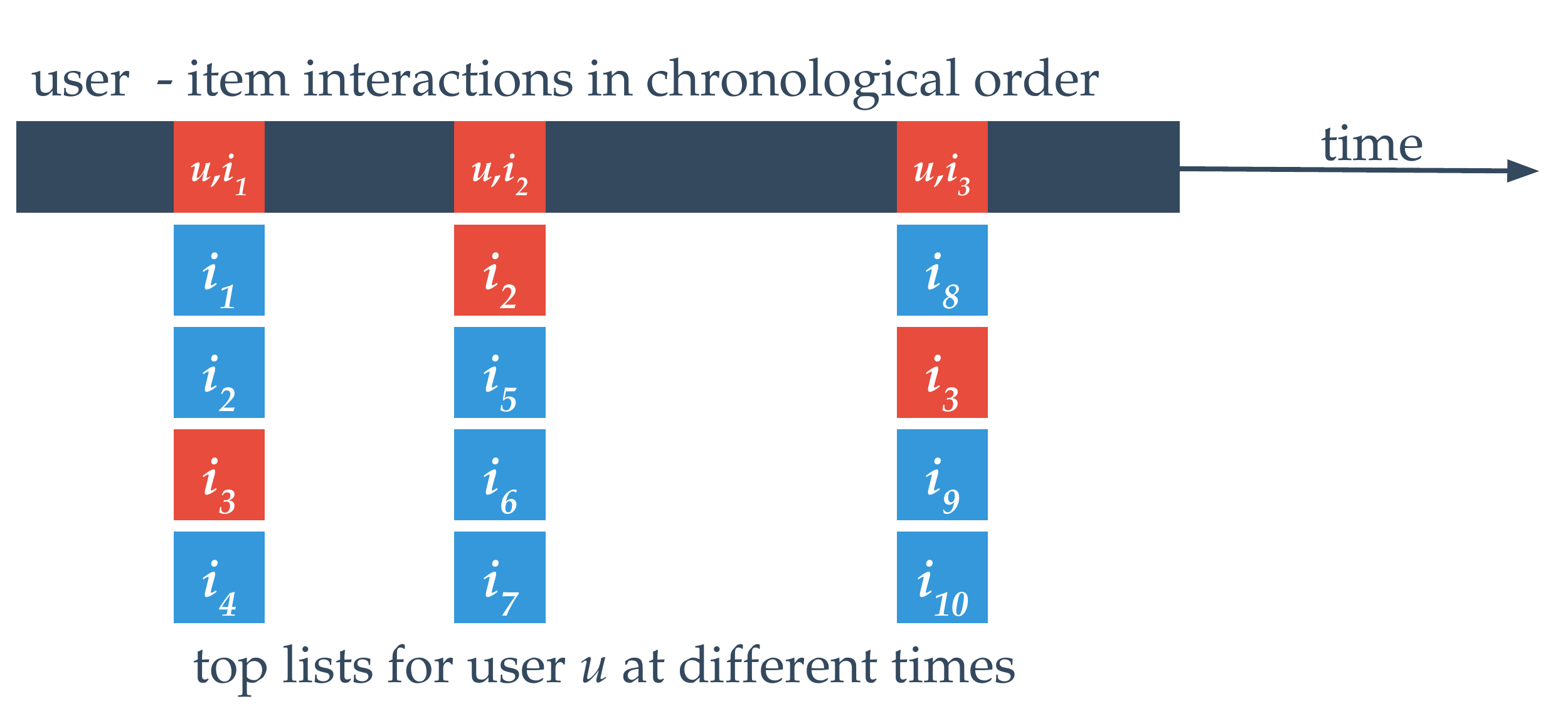}
\caption{Prequential evaluation of the online ranking prediction problem.}
\label{fig:online}
\end{figure}
\begin{enumerate}
\item We query the recommender for a top-$k$ recommendation for the active user.
\item We evaluate the list in question against the single relevant item that the user interacted with.
\item We allow the recommender to train on the revealed user-item interaction.
\end{enumerate}
Since we can potentially re-train the model after every new event, the recommendation for the same user may be very different even at close points in time, as seen in Fig.~\ref{fig:online}.
The standard recommender evaluation settings used in research cannot be applied, since there is always only a single relevant item in the ground truth.

In one of the possible recommender evaluation settings, the \textbf{rating prediction} problem, which is popular in research, we consider a user $u$ and an item $i$.  The actual user preference in connection with the item is expressed as a value $r_{u i}$, for which the system returns a prediction $\hat{r}_{ui}$.  This explicit rating can be a scale such as one to five stars for a Netflix movie, while implicit rating can be the duration of viewing a Web page in seconds.  Implicit rating is binary when the only information is whether the user interacted with the item (clicked, viewed, purchased) or not.  Depending on whether $r_{u i}$ is binary or scale, the same prequential metrics, such as error rate or MSE, can be applied as for classification or regression.  For example, in the Netflix Prize competition, the target was the square root of MSE between the predicted and actual ratings, see Equation~(\ref{eq:mse}).

Another possible way to evaluate recommenders is \textbf{ranking prediction}, where performance metrics depend on the list of displayed items.
We note that given rating prediction values $\hat{r}_{ui}$ for all $i$, in theory, ranking prediction can be solved by sorting the relevance score of all items.  For certain models, heuristics to speed up the selection of the highest values of $\hat{r}_{ui}$ by candidate preselection exist~\cite{teflioudi2015lemp}.

To evaluate ranking prediction, we have to take into consideration two issues that do not exist for classifier evaluation.
In the case of prequential evaluation, as shown in Fig.~\ref{fig:online}, the list for user $u$ may change potentially after every interaction with $u$.  As soon as $u$ provides feedback for certain item $i$, we can change model parameters and the set of displayed items may change completely.  Most of the batch ranking quality measures focus on the set of items consumed by the same user, under the assumption that the user is exposed to the same list of items throughout the evaluation.  As this assumption does not hold, we need measures for individual user-item interactions.

Another issue regarding ranking prediction evaluation lies in a potential user-system interaction that affects quality scores.  Typically, the set of items is very large, and users are only exposed to a relatively small subset, which is usually provided by the system.  The form of user feedback is usually a click on one or more of these items, which can be evaluated by computing the clickthrough rate.  Since users cannot give feedback on items outside the list, the fair comparison of two algorithms that present different sets for the user can only be possible by relying on live user interaction.
This fact is known by practitioners, who use \textbf{A/B testing} to compare the performance of different systems.  In A/B testing, the live set of users is divided into groups that are exposed to the results of the different systems.


Most traditional ranking prediction metrics, to a certain level, rely on the assumption that the same user is exposed to the same list of items, and hence the interactions of the same user can be considered to be the unit for evaluation.
For online evaluation, as noted in~\cite{palovics2014exploiting}, the unit of evaluation will be a single interaction, which usually contains a single relevant item.  Based on this modification, most batch metrics apply in online learning evaluation as well.  
Note that the metrics below apply not just in A/B testing, but also in experiments with frozen data, where user feedback is not necessarily available for the items returned by a given algorithm.  For example, if the item consumed by the user in the frozen data is not returned by the algorithm, the observed relevance will be 0, which may not be the case if the same algorithm is applied in an A/B test.
Note that attempts to evaluate research results by A/B testing have been made in the information retrieval community~\cite{balog2014head}; however, designing and implementing such experiments is cumbersome.

Next, we list several metrics for the quality of the ordered top-$K$ list of items $L = \lbrace {i_1, i_2, ..., i_K}\rbrace$ against the items $E$ consumed by the user.
We will also explain how online evaluation metrics differ from their batch counterparts.  For the discussion, we mostly follow~\cite{palovics2014exploiting}.

Clickthrough rate is commonly used in the practice of recommender evaluation.  It is defined as the ratio of clicks received for $L$:
\begin{equation}
  \mbox{Clickthrough@K} =
  \cases{1 & if  $E \cap L \not= \emptyset$; \cr
    0 & otherwise.}
\end{equation}
For precision and recall, similar to Clickthrough, the actual order within $L$ is unimportant:
\begin{equation}
        \mbox{Precision@K} = \frac{|E \cap L|}{K}, \quad  \mbox{Recall@K} = \frac{|E \cap L|}{|E|}.
\end{equation}
For batch evaluation, $E$ is the entire set of items with positive feedback from a given user who is exposed to the same $L$ for each interaction.  The overall batch system performance can be evaluated by averaging precision and recall over the set of users.
For online evaluation, typically $|E|=1$, where Precision@K is 0 or $1/K$ and Recall@K is 0 or 1 depending on whether the actual item in $E$ is listed in $L$ or not.  Precision and recall are hence identical to clickthrough, up to a constant.  As a consequence, the properties of online precision and recall are very different from their batch counterparts.
The main reason for the difference lies in the averaging procedure of prequential evaluation: We cannot merge the events of the same user, instead, we average over the set of individual interactions.

Measures that consider the position of the relevant item $i$ in $L$ can give more refined performance indication.
The first example is reciprocal rank:
\begin{equation}
          RR@K =
          \cases{
        0 & if rank${}(i) > K$; \cr
        \frac{\displaystyle 1}{\displaystyle \mbox{rank}(i)} & otherwise.}
\end{equation}
Discounted cumulative gain (DCG) is defined similarly, as
\begin{equation}
        \mbox{DCG@K} = \displaystyle\sum_{k=1}^K \frac{\mbox{rel}(i_k)}{\log_2(1 + k)}
        \label{eq:dcg}
\end{equation}
where $\mbox{rel}(i_k)$ indicates the relevance of the i-th item in the list.
For the implicit task, relevance is 1 if the user interacted with the item in the evaluation set, 0 otherwise.
For batch evaluation, we can consider all interactions of the same user as one unit. If we define iDCG@K, the ideal maximum possible value of DCG@K for the given user, we can obtain nDCG@K, the normalized version of DCG@K, as
\begin{equation}
\mbox{nDCG@K} = \frac{\mbox{DCG@K}}{\mbox{iDCG@K}}.
\label{eq:ndcg}
\end{equation}
Note that for online learning, there is only one relevant item, hence iDCG${}=1$. For emphasis, we usually use the name nDCG for batch and DCG for online evaluation.

\subsection{Session-Based Recommendation}

Previous items in user sessions constitute a very important context~\cite{hidasi2016general}.  In e-commerce, the same user may return next time with a completely different intent and may want to see a product category completely different from the previous session.
Algorithms that rely on recent interactions of the same user are called \textbf{session-based item-to-item} recommenders.  The user session is special context, and it is the only information available for an item-to-item recommender.
In fact, several practitioners~\cite{koenigstein2013towards,pilaszy2015neighbor} argue that most of the recommendation tasks they face are without sufficient past user history. For example, users are often reluctant to create logins and prefer to browse anonymously.  Moreover, they purchase certain types of goods (for example, expensive electronics) so rarely that their previous purchases will be insufficient to create a meaningful user profile.
Whenever a long history of previous activities or purchases by the user is not available, recommenders may  propose items that are similar to the most recent ones viewed in the actual user session.

Session-based recommendation can be served by very simple algorithms, most of which are inherently online.  A comparison of the most important such online algorithms in terms of performance is available in~\cite{frigo2017online}.
Data stream processing algorithms can retain items from the most recently started sessions as long as they fit in their memory.  Recommendation is based on the recent items viewed by the user in the actual shopping session.
For example, we can record how often users visited item $i$ after visiting another item $j$. Since fast update to transition frequencies is usually possible, the method is online.

In an even simpler algorithm that is not strictly session-based, we recommend the most popular recent items. This method can be considered batch or online depending on the granularity of the item frequency measurement update.  
Both algorithms can be personalized if we consider the frequency of past events involving the user.  If items are arranged hierarchically (for example, music tracks by artist and genre), personal popularity and personal session data can involve the frequency of the artists or genres for recommending tracks.  More session-based algorithms are described in~\cite{koenigstein2013towards}.


\subsection{Online Matrix Factorization}
\label{sec:mf}

\begin{figure}
  \sidecaption
\includegraphics[width=7cm]{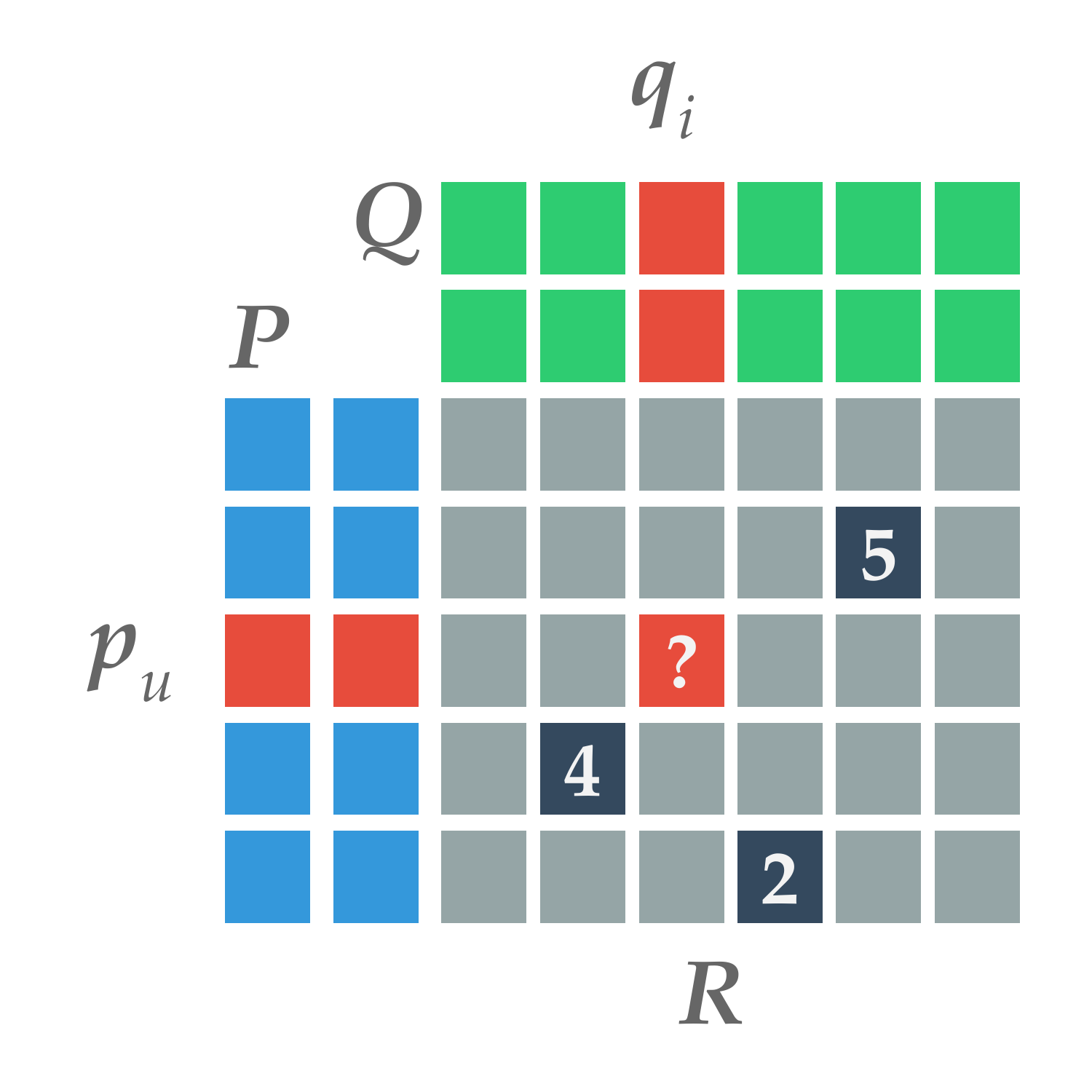}
\caption{Utility matrix $R$ and the matrix factorization model built from matrices $P$ and $Q$.}
\label{fig:utility-matrix}
\end{figure}

Most nontrivial online recommender algorithms are based on matrix factorization~\cite{koren2009matrix}, a popular class of collaborative filtering methods.
Given the user-item utility matrix $R$ shown in Fig.~\ref{fig:utility-matrix}, we model $R$ by decomposing it into the two dense matrices $P$ and $Q$.
For a given user $u$, the corresponding row in $P$ is user vector $p_u$.
Similarly, for item $i$, the corresponding column of $Q$ is item vector $q_i$.
The predicted relevance of item $i$ for user $u$ is then
\begin{equation}
  \hat{r}_{ui} = p_u q_i^T.
  \label{eq:mf}
\end{equation}
Note that we can extend the above model by scalar terms that describe the biased behavior of the users and the items~\cite{koren2009matrix}.

One possibility to train model parameter matrices $P$ and $Q$ is by gradient descent~\cite{koren2009matrix,funk2006netflix}, which can be applied to online learning as well \cite{palovics2014exploiting}.
For a set of interactions $E$, we optimize  equation (\ref{eq:mse}) for MSE as target function.
In one step of gradient descent, we fit $P$ and $Q$ in equation~(\ref{eq:mf}) to one of the ratings in $E$.  Unlike in batch training, where we can use the ratings several times in any order, in online learning, we have the most recent single item in $E$.  In other words, in online gradient descent, we fit the model to the events one by one as they arrive in the data stream.

For a given (explicit or implicit) rating $r_{ui}$, the steps of gradient descent are as follows.  First, we compute the gradient of objective function $F$ with respect to the model parameters:
\begin{equation}
        \frac{\partial F}{\partial p_u} =   - 2 (r_{ui} - \hat{r}_{ui}) q_i, \quad
        \frac{\partial F}{\partial q_i} =   - 2 (r_{ui} - \hat{r}_{ui}) p_u.
\end{equation}
Next, we update the model parameters in opposite direction of the gradient, proportionally to learning rate $\eta$, as
\begin{eqnarray*}
        p_u &\leftarrow&  \eta (r_{ui} - \hat{r}_{ui}) q_i, \\
        q_i &\leftarrow&  \eta (r_{ui} - \hat{r}_{ui}) p_u.
\end{eqnarray*}
Overfitting is usually avoided by adding a regularization term in the objective function~\cite{koren2009matrix}.

In the case of implicit feedback, the known part of the utility matrix only contains elements with positive feedback.
To fit a model, one requires negative feedback for training as well. Usually, such elements are selected by sampling from those that the user has not interacted with before~\cite{rendle2014improving}.
We can also introduce confidence values for ratings and consider lower confidence for the artificial negative events~\cite{hu2008collaborative}.

Gradient descent can also be used in a mix of batch and online learning, for example, training batch models from scratch periodically, and continuing the training with online learning.
We can also treat users and items differently, for example, updating user vectors more dynamically than item vectors, as first suggested by~\cite{takacs2009scalable}.

Another use of online gradient descent is to combine different recommendation models~\cite{palovics2014exploiting}.  We can express the final prediction as the linear combination of the models in the ensemble whose parameters are the linear coefficients and the individual model parameters. In~\cite{palovics2014exploiting}, two online gradient descent methods are described with regards to whether the derivative of the individual models is available, where all parameters can be trained through the derivative of the final model or otherwise by learning the coefficients and the individual models separately.

\subsection{Variants of Matrix Factorization}

Several variants of matrix factorization that can be trained by gradient descent both for batch and online learning tasks have been proposed.
Bayesian Personalized Ranking~\cite{rendle2009bpr} has top list quality as target instead of MSE.
In asymmetric matrix factorization~\cite{patarekamf}, we model the user by the sum of the item vectors the user rated in the past.

Recently, various factorization models have been developed that incorporate context information~\cite{hidasi2016general}.
Context data can be modeled by introducing data tensor $D$ instead of the rating matrix $R$.
In a simplest case, the data includes a single piece of additional context information~\cite{rendle2010pairwise}: for example, music tracks can have artist as context.

Alternating Least Squares~\cite{koren2009matrix,pilaszy2010fast} (ALS) is another optimization method for matrix factorization models, in which for a fixed $Q$, we compute the optimal $P$, then for a fixed $P$, the optimal $Q$, repeatedly until certain stopping criteria are met.
Hidasi et al.~\cite{hidasi2012fast,hidasi2014factorization,hidasi2016general} introduced several variants of ALS-based optimization schemes to incorporate context information.
By incremental updating, ALS can also be used for online learning~\cite{he2016fast}.

\subsection{Summary of Recommendation by Online Learning}

Recommendation differs from classification in that in recommendation, there are two types of objects, users, and items, and a prediction has to be made for their interaction.  A practical recommender system  displays a ranked list of a few items for which the user can give feedback.  In an online learning system, the list shown to the same user at different times may change completely for two reasons.  First, as in the prequential classifier training and evaluation setting, the list of recommendations may change because the model changes.  Second, the user feedback we use for evaluation depends on the actual state of the model, since the user may have no means to express interest in an item not displayed.  Hence for online learning evaluation, metrics that involve the notion of a volatile list have to be used.

Online learning is very powerful for recommender systems due to their advantage of having much more emphasis on recent events.  For example, if we update models immediately for newly emerged users and items, trends are immediately detected.
The power of online learning for recommendation may also be the result of updating user models with emphasis on recent events, which may be part of the current user session.
User session is a highly relevant context for recommendation and most session-based methods are inherently online.

\section{Additional Topics}
\label{sect:additional}

In this final section, we give a brief overview of two additional topics, both of which are extensively covered in recent surveys.  In Section~\ref{sect:unsupervised}, we describe unsupervised data mining methods, including clustering, frequent itemset mining, dimensionality reduction, and topic modeling.  In Section~\ref{sect:drift}, we describe the notion of the dataset drift, or in other terms, concept drift, and list the most important drift adapting methods.  We only discuss representative results in these areas.


\subsection{Unsupervised Data Mining}
\label{sect:unsupervised}

The most prominent class of unsupervised learning methods is \textbf{clustering} where instances have to be distributed into a finite set of clusters such that instances within the cluster are more similar to each other than to others in different clusters~\cite{pang2006introduction}.  Batch clustering algorithms have been both studied and employed as data analysis tools for decades~\cite{jain1999data,xu2008clustering}.  One frequently applied clustering method is \textbf{k-means}~\cite{hartigan1975clustering} where cluster center selection and assignment to nearest centers are iteratively performed until convergence.  Another is \textbf{DBSCAN}~\cite{ester1996density}, a density-based method that groups points that are closely packed together.

Online clustering algorithms are surveyed among other places in~\cite{mahdiraji2009clustering,kavitha2010clustering,aggarwal2013survey,silva2013data}.  The majority of the most relevant methods are data stream versions of k-means or its variants such as k-medians~\cite{zhang1996birch,bradley1998scaling,farnstrom2000scalability,o2002streaming,guha2003clustering,aggarwal2003framework,zhou2008tracking,gama2011clustering,kranen2011clustree,ackermann2012streamkm}.  Another set of results describes the data stream implementation of DBSCAN~\cite{cao2006density,chen2007density,kranen2011clustree}.  Finally, an online hierarchical clustering algorithm that maintains similarity measures and hierarchically merges closest clusters is described in~\cite{rodrigues2006odac}.

Finding \textbf{frequent itemsets}~\cite{Agrawal1993-ais} is another central unsupervised data mining task, both static and streaming. In brief, for a table of transactions and items, the task is to find  all subsets of items that occur together in transactions with at least a prescribed frequency.  Several variants of the task are described in~\cite{aggarwal2014frequent}.  Online frequent itemset mining algorithms are surveyed in~\cite{cheng2008survey}. Algorithms based on counts of all past data in the stream~\cite{chang2003finding,giannella2003mining,li2004efficient,yu2004false,lee2005finding} are also called landmark window based approaches.  In some of these algorithms, time adaptivity is achieved by placing more importance on recent items~\cite{chang2003finding,giannella2003mining,lee2005finding}.
Sliding window based approaches~\cite{chang2003estwin,chi2006catch,chang2006finding,song2007claim,cheng2008maintaining,li2009incremental,yen2011fast,calders2014mining} are particularly suitable for processing data with concept drift.  For a comparative overview, see, for example, how MOA's algorithm was selected~\cite{quadrana2015efficient}.
Note that a special subtask, finding frequent items in data streams, is already challenging and requires approximate data structures~\cite{charikar2004finding}.

\textbf{Principal component analysis (PCA)} is a powerful tool for dimensionality reduction~\cite{jolliffe1986principal} based on matrix factorization.  Online variants are based on ideas to incrementally update the matrix decomposition~\cite{bunch1978updating,hall2000merging,brand2002incremental}.  The first PCA algorithms suitable for online learning are based on neural networks~\cite{oja1982simplified,sanger1989optimal,oja1992principal}.  Similar to the linear models in Section~\ref{sect:linear}, PCA can also apply the kernel trick to involve nonlinear modeling~\cite{scholkopf1998nonlinear}.  Iterative \textbf{kernel PCA} is described in~\cite{kim2005iterative,gunter2007fast} and online kernel PCA in~\cite{honeine2012online}.  We note that for nearest neighbor search in the low-dimensional space provided by PCA, the heuristics for selecting large inner products is applicable~\cite{teflioudi2015lemp}.

\textbf{Probabilistic topic modeling} fits complex hierarchical Bayesian models to large document collections. A topic model reveals latent semantic structure that can be used for many applications.  While PCA-like models can also be used for latent semantic analysis~\cite{deerwester90indexing}, recently the so-called \textbf{Latent Dirichlet Allocation~(LDA)}~\cite{blei2003latent} has gained popularity.
Most topic model parameters can only be inferred based on Markov Chain Monte Carlo sampling, a method difficult to implement for online learning.  LDA inference is possible based on either online Gibbs sampling~\cite{song2005modeling,canini2009online} or online stochastic optimization with a natural gradient step~\cite{hoffman2010online}.
Several online LDA variants are described in~\cite{smola2010architecture,ho2013more,li2014scaling,yuan2015lightlda,yu2015scalable,jagerman2017computing}.

\subsection{Concept Drift and Adaptive Learning}
\label{sect:drift}

In dynamically changing and nonstationary environments, we often observe concept drift as the result of data distribution change over time.
The phenomenon and mitigation of concept (or dataset) drift for online learning are surveyed in several articles~\cite{widmer1996learning,tsymbal2004problem,quionero2009dataset,zliobaite2012next,gama2014survey}.
The area of transfer learning where the (batch) training and the test sets are different~\cite{pan2010survey}, is closely related to concept drift~\cite{storkey2009training}, but more difficult in the sense that adaptation by learning part of the new data is not possible.

Adaptive learning refers to the technique of updating predictive models online to react to concept drifts.  One of the earliest active learning systems is STAGGER~\cite{schlimmer1986incremental}. In~\cite{zliobaite2009learning}, the main steps of online adaptive learning are summarized as (1) making assumptions about future distribution, (2) identifying change patterns, (3) designing mechanisms to make the learner adaptive, and (4) parameterizing the model at every time step.

A comprehensive categorization of concept drift adaptation techniques is found in~\cite{gama2014survey}.  Online learning algorithms can naturally adapt to evolving distributions.  However, adaptation happens only as the old concepts are diluted due to the new incoming data, which is more suitable for gradual changes~\cite{littlestone1988learning,domingos2000mining}.
For sudden changes, algorithms that maintain a sliding window of the last seen instances perform better~\cite{widmer1996learning,gama2004learning,kuncheva2009window}.
Another option is to include explicit forgetting mechanisms~\cite{koychev2000gradual,klinkenberg2004learning,elwell2009incremental}.
The most important distinction is whether changes are explicitly or implicitly detected:
\textbf{Trigger-based} methods aim at detecting when concept drift occurs to build a new model from scratch~\cite{gama2004learning}.
\textbf{Evolving learners}, by contrast, do not aim to detect changes but rather maintain the most accurate models at each time step.  Evolving learners are method-specific, most of them based on ensemble methods~\cite{wang2003mining,kolter2003dynamic}.

A few papers~\cite{minku2010impact,moreno2012unifying} give overviews of different types of environmental changes and concept drifts based on speed, recurrence, and severity.  Drift can happen gradually or suddenly, in isolation, in tendencies or seasonally, predictably or unpredictably, and its effect on classifier performance may or may not be severe.   In~\cite{schlimmer1986incremental,gama2004learning}, several artificial data sets with different drift concepts, sudden or abrupt, and gradual changes are described.

A large variety of single classifier and ensemble models capable of handling concept drift are described in~\cite{tsymbal2004problem}.  Perhaps the majority of the results consider tree-based methods~\cite{alberg2012knowledge}.
For example, concept drift adaptive online decision trees based on a statistical change detector that works on sliding windows are described in~\cite{bifet2009adaptive,bifet2010adaptive}.
More examples include Bayesian models~\cite{gama2003accurate,bach2010bayesian}, neural networks~\cite{gama2007stream,leite2013evolving}, and SVM~\cite{syed1999handling,klinkenberg2000detecting}.
Concept drift adaptation methods exist for clustering~\cite{rodrigues2006odac,silva2013data}.  Sliding window based data stream frequent itemset mining is also adaptive~\cite{quadrana2015efficient}.
Some of the results do not follow the data stream computational model but rather use computational resources with little restriction.  One class of such methods are incremental algorithms with partial memory~\cite{maloof2004incremental}.
We also note that there is a MOA-based software system for concept drift detection~\cite{bifet2013cd}.

\bibliographystyle{spmpsci}


\end{document}